# Effect of Singular Value Decomposition Algorithms on Removing Injection Variability in 2D Quantitative Angiography of Intracranial Aneurysms

Parmita Mondal[1,2], Swetadri Vasan Setlur Nagesh[2,4], Sam Sommers-Thaler[1,2], Allison Shields[1,2], Mohammad Mahdi Shiraz Bhurwani[3], Kyle A Williams[1,2], Ammad Baig[1,2,4], Kenneth Snyder[1,2,4], Adnan H Siddiqui[1,2,4], Elad Levy[1,2,4], Ciprian N Ionita[1,2,3]

[1]Department of Biomedical Engineering, University at Buffalo, Buffalo, NY 14260
[2]Canon Stroke and Vascular Research Center, Buffalo, NY 14203
[3]Quantitative Angiographic Systems. Artificial Intelligence, Buffalo, NY 14203
[4]University at Buffalo Neurosurgery, Inc, Buffalo, NY 14203

## Abstract

**Background**: Intraoperative 2D quantitative angiography (QA) for intracranial aneurysms (IAs) has accuracy challenges due to the variability of hand injections. Despite the success of singular value decomposition (SVD) algorithms in reducing biases in computed tomography perfusion (CTP), their application in 2D QA has not been extensively explored. This study seeks to bridge this gap by investigating the potential of SVD-based deconvolution methods in 2D QA, particularly in addressing the variability of injection durations.

**Purpose:** Building on the identified limitations in QA, the study aims to adapt SVD-based deconvolution techniques from CTP to QA for IAs. This adaptation seeks to capitalize on the high temporal resolution of QA, despite its two-dimensional nature, to enhance the consistency and accuracy of hemodynamic parameter assessment. The goal is to develop a method that can reliably assess hemodynamic conditions in IAs, independent of injection variables, for improved neurovascular diagnostics.

**Materials and Methods:** The study included three internal carotid aneurysm (ICA) cases. Virtual angiograms were generated using Computational Fluid Dynamics (CFD) for three physiologically relevant inlet velocities to simulate contrast media injection durations. Time-density curves (TDCs) were produced for both the inlet and aneurysm dome. Various SVD variants, including standard SVD (sSVD) with and without classical Tikhonov regularization, block-circulant SVD (bSVD), and oscillation index SVD (oSVD), were applied to virtual angiograms. The method was applied on virtual angiograms to recover the aneurysmal dome impulse response function (IRF) and extract flow related parameters such as Peak Height $PH_{IRF}$, Area Under the Curve $AUC_{IRF}$, and Mean transit time MTT. Next, correlations between QA parameters, injection duration and inlet velocity were assessed for unconvolved and deconvolved data for all SVD methods.

**Results:** The different SVD algorithm variants showed strong correlations between flow and deconvolution-adjusted QA parameters. Furthermore, we found that SVD can effectively reduce QA parameter variability across various injection durations, enhancing the potential of QA analysis parameters in neurovascular disease diagnosis and treatment.

**Conclusions:** Implementing SVD-based deconvolution techniques in QA analysis can enhance the precision and reliability of neurovascular diagnostics by effectively reducing the impact of injection duration on hemodynamic parameters.

## 1. Introduction

An intracranial aneurysm (IA) is characterized as a localized and abnormal dilation of a blood vessel, typically an artery, that emerges from the weakening of the vessel wall. With a global prevalence rate of 3.2% and an equal gender ratio, cerebral aneurysms predominantly afflict individuals around the median age of 50. The rupture of these aneurysms, leading to subarachnoid hemorrhage (SAH), occurs at a rate of approximately 10 per 100,000 people. The overall mortality due to aneurysmal SAH is 0.4 to 0.6% of all-cause deaths, with an approximate 20% mortality and an additional 30 to 40% morbidity in patients with known rupture[1]. Hemodynamics plays a critical role in determining the risk of aneurysm rupture, as well as in the growth and post-treatment progression of an aneurysm[2]. Understanding these hemodynamic patterns is pivotal for selecting appropriate treatment strategies.

Altering this hemodynamics is central to the management of neurovascular aneurysms. This understanding directly informs the use of specific endovascular techniques. Interventionalists employ endovascular techniques, which are less invasive compared to open surgery, to access the treatment site. These procedures often involve devices like coils and flow diverters. Coils induce clotting to isolate the aneurysm, while flow diverters redirect blood flow away from the aneurysm, promoting vessel healing. However, due to the aneurysm's remote location and the intricate, tortuous nature of the vasculature, real-time quantitative assessment of blood flow within IAs during intervention remains a challenge. Consequently, interventionalists rely on a qualitative evaluation based on the intra-arterial flow of the contrast agent via Digital Subtraction Angiography (DSA). Despite its limitations associated with its two-dimensional aspect and hand injection induced variability, DSA remains the primary imaging modality during these procedures, providing critical structural visualization of blood vessels and aneurysms.

To address these limitations, advanced techniques such as Quantitative Angiography (QA) have been developed. QA allows for a more detailed analysis of blood flow dynamics. Leveraging the high temporal and spatial resolution of DSA, QA becomes an effective tool for enhancing the precision of measurements and analysis of contrast/tracer flow parameters during interventions.[3,4] The process begins with administering contrast via a catheter in the proximity of the arterial segment of interest. Subsequently, QA utilizes this contrast flow, which is indicative of the local hemodynamic conditions, to extract valuable imaging biomarkers. This is achieved by synthesizing a time density curve (TDC) at every pixel or in a region of interest manually placed by the user, using information from each frame.[5] The TDC is a critical tool in QA, as it represents the temporal changes in contrast density at specific locations within the blood vessels, providing insights into the flow dynamics that are correlated with local hemodynamics.**[6]** Through this approach, QA offers a more nuanced estimation of the hemodynamic environment within aneurysms, bridging the gaps left by traditional DSA and advancing our understanding of aneurysm behavior and treatment strategies.

As noted previously,[7-9] one of the significant challenges in QA is managing the variability caused by hand or mechanical injection methods. This variability introduces uncertainty that can notably impact the precision of hemodynamic estimations. Such challenges are not new to angiography and have been extensively investigated in computed tomography perfusion (CTP), where arterial inlet TDC, known as the arterial input function (AIF), is deconvolved from tissue TDCs at every voxel to mitigate these effects.[10-12] The deconvolution is performed using standard or modified Singular Value Decomposition (SVD) algorithms. While these techniques have shown varying levels of success at CTP, their application to 2D

QA has not been fully explored. This gap presents an opportunity to adapt and test these methods in the context of QA, particularly for aneurysm hemodynamic research.

Deconvolution is a mathematical process used to reduce the AIF effects from time density curves. This separation is essential for more accurately estimating QA parameters, thereby countering the effects of injection duration variability. Prior studies have indicated that truncated SVD can be effective in QA across different injection durations. Building on this foundation, our study aims to investigate the extension of this SVD-based deconvolution techniques across the spectrum of QA. By applying this method to virtual angiograms, we seek to determine whether such deconvolution techniques, successful in CTP, can be effectively adapted to enhance the accuracy of hemodynamic assessments in QA for aneurysm research.

In summary, our objective is to investigate the application of SVD-based deconvolution techniques, well-established in CTP, in the realm of QA. We propose to perform a comprehensive in silico study where hemodynamic parameters and virtual injection parameters are controlled with high accuracy in patient specific IA geometries. In doing so, we aim to establish a more accurate and reliable framework for estimating hemodynamics in IAs during interventions, ultimately contributing to enhanced diagnostic precision and tailored treatment strategies in neurovascular care.

## 2. Materials and Methods

**2.1 Theoretical background**

In this work, we employ tracer-based methodologies to estimate hemodynamic conditions. Tracers, like iodinated contrast agents, are used in medical imaging to track fluid movement and transport quantities. The modified Fick principle, relating blood flow to tracer exchange, underpins our approach[13]. It equates arterial inflow to venous outflow in cerebral tissue (Equation 1). In medical imaging, this principle aids in understanding the pharmacokinetics of contrast agents and how injection duration affects the contrast bolus' characteristics.

$$Q(t) = F \cdot \int_0^t C_a(t - t') IRF(t') \cdot dt' \tag{1}$$

where: $Q(T)$ is mass of contrast in the mass of brain tissue at time T, in our case the aneurysm dome, $t'$ is variable of integration, $t$ is time, $F$ is perfusion in units of ml. (min)$^{-1}$ (100g)$^{-1}$, $C_a$ is the contrast concentration in arterial inlet also referred as the Arterial Input Function (AIF), *IRF* is called the impulse response function (IRF) – the tissue residue function corresponding to the instantaneous deposition of unit mass of contrast

The first step in the development of a deconvolution algorithm is to discretize the above equation and rewrite it in matrix form:

$$Q = F \cdot \Delta t \cdot C_a \cdot IRF \tag{2}$$

In the field CTP, deconvolution techniques, particularly those based on SVD, are commonly employed to reduce the dependency of QA parameters on contrast injection parameters.[14] This reduction is crucial for generating more standardized and reliable perfusion maps. SVD, a fundamental technique in linear algebra, decomposes a matrix into three constituent matrices (U, S, V$^T$) as shown in Equation 2. Its application is advantageous in our context due to its ability to manage the ill-posed nature of inverse problems commonly encountered in medical imaging data analysis. [15] [16]

$$C_a = AIF = U \cdot S \cdot V^T \quad (3)$$

where: U is left singular vectors, S is diagonal singular vectors, $V^T$ is the transform of right singular vectors.

From equations 2 and 3, an estimate of the flow scaled IRF $\widetilde{R}$ than becomes:

$$F \cdot IRF = \widetilde{R} = \frac{1}{\Delta t} \cdot V^T \cdot S^{-1} \cdot U \cdot Q \quad (4)$$

In this study, we evaluate the appropriateness of various SVD formalisms for use in 2D QA for three IAs shown in Figure 1. Figure 1 shows each of the three ICA cases, taken into account for this study. M1, M2 and M3 are three aneurysm geometries, chosen for this study. For acquiring the AIF– essentially the TDC for the inlet, we place a region of interest on the inlet artery in a segment exhibiting minimal tortuosity and foreshortening. $Q$ represents in actuality the TDC recorded from the ROI placed over the aneurysm dome. The intra-aneurysmal dome flow scaled IRF, i.e., $\widetilde{R}$, is derived from Equation 4, and it indicates the fraction of contrast media that remains in the blood IA dome as the time evolves.

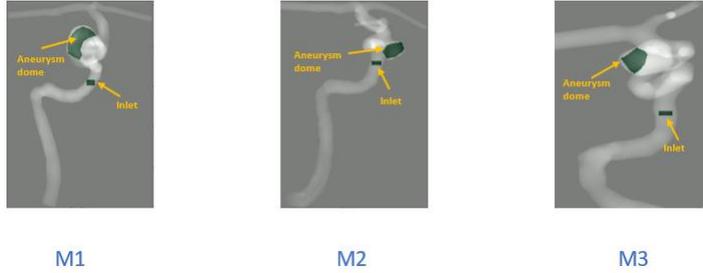

**Figure 1.** Each of the three aneurysm geometries are shown., where yellow arrows show the aneurysm dome and inlet ROI. M1, M2 and M3 are three aneurysm geometries, chosen for this study.

The primary challenge in performing these calculations lies in the accurate execution of numerical methods for Equation 3 under conditions of noise, sparse data, and other image-related artifacts. Over the past decade, various numerical methods have been developed to address these challenges. The variants explored include standard SVD with classical Tikhonov regularization (sSVD)[17], block-circulant SVD (bSVD)[18], and oscillation index SVD (oSVD).[19] These methods aim to optimize the balance between fidelity to the measured data and regularization to mitigate the effects of noise and ill-conditioning and will be briefly described in the context of 2D-QA.

In the sSVD approach, the $C_a$ – the **AIF** is transformed into a Toeplitz matrix This matrix serves as the basis for applying the sSVD technique, facilitating the decomposition of the $C_a$ without implying exclusivity or necessity of this method for the analysis:

$$C_a^{toep} = \begin{bmatrix} C_a(t_1) & 0 & \cdots & 0 \\ C_a(t_2) & C_a(t_1) & \cdots & 0 \\ \vdots & \vdots & \vdots & \vdots \\ C_a(t_n) & C_a(t_{n-1}) & \cdots & C_a(t_{n-2}) \end{bmatrix} \quad (5)$$

After transforming the AIF into a Toeplitz matrix, we proceed with its SVD, breaking it down into the matrices U, V, and S, as delineated in Equation 3. This decomposition utilizes Python's SciPy library for linear algebra, ensuring a robust analytical framework. To address the challenge of imaging noise, we apply a truncation strategy to the S matrix, for example an optimal truncation (SVD$_{trunc}$) of 10%. This means only retaining singular values that are greater than 10% of the maximum singular value in S, setting all others to zero, thus enhancing data clarity. Furthermore, to counteract potential overfitting, we employ Tikhonov regularization (TR) on the S matrix, a method also known as Ridge Regression.[17] The regularization process is integrated with the SVD framework using the following formula:

$$TR = \frac{S}{S^2 + \lambda^2} \quad (6)$$

Where: S= diagonal singular vectors, λ is the regularization factor. Thus, each value in the S matrix is calculated by dividing each value in the diagonal S matrix, by the sum of its square and square of lambda (λ). λ represents the regularization factor, which is 0.1 for this study.

In the second case, we study implementation of the bSVD. Unlike the first method, bSVD transforms the arterial input function (AIF) into a block-circulant matrix. [18] This transformation enables a different approach to singular value decomposition, designed to address imaging artifacts more effectively, particularly those from patient motion. Based on the $C_a$ values the AIF matrix can be rewritten as:

$$T_{bl-cir} = \begin{bmatrix} C_a(t_1) & C_a(t_n) & \ldots & C_a(t_3) & C_a(t_2) \\ C_a(t_2) & C_a(t_1) & C_a(t_n) & \ldots & C_a(t_3) \\ \vdots & C_a(t_2) & C_a(t_1) & \ddots & \vdots \\ C_a(t_{n-1}) & \ddots & \ddots & \ddots & C_a(t_n) \\ C_a(t_n) & C_a(t_{n-1}) & \ldots & C_a(t_2) & C_a(t_1) \end{bmatrix} \quad (7)$$

Like in the previous method, the choice of threshold for truncating singular values in bSVD is determined by the need to balance noise suppression with the preservation of the integrity of the impulse response function (IRF), highlighting a methodological contrast to Tikhonov regularization.

In the third methodological approach of our study, we explore the oscillation-index SVD (oSVD) technique. oSVD involves transforming the inlet function into a block-circulant matrix, subsequently decomposed into U, S, and V matrices. The primary objective of oSVD is to address and minimize oscillations at the tail end of the impulse response function (IRF) by iteratively adjusting the SVD truncation threshold for the diagonal S matrix[20]. This iterative process entails calculating the oscillation index (OI) after each extraction of IRF using Equation 8:

$$OI = \frac{1}{L} \frac{1}{f_{max}} \frac{1}{\triangle t} \left( \sum_{k=2}^{L-1} |f(k) - 2f(k-1) + f(k-2)| \right) \quad (8)$$

where f is the amplitude of IRF, $f_{max}$ is the maximum amplitude of IRF, Δt is the time step and L is the number of sample points. The IRF is generated by adjusting the SVDtrunc so that we observe minimum oscillations at the tail of IFR. If the OI is found to be excessively high, indicating significant oscillations, the SVD truncation threshold is modified and the process is repeated to ensure minimal oscillations, thereby improving the accuracy of subsequent IRF-derived parameters. The rationale behind the development of oSVD and its iterative approach stems from the method's ability to refine the deconvolution process in medical imaging, particularly in cerebral perfusion studies. By specifically targeting the reduction of non-physiological oscillations in the IRF, oSVD aims to enhance the precision of perfusion quantification. This methodological innovation addresses a critical challenge in quantifying cerebral perfusion, where non physiological oscillations in the IRF and underestimation of perfusion can significantly impact the accuracy of derived hemodynamic parameters.

In the context of SVD-based medical imaging analysis, the IRF provides a mathematical representation of how contrast media disperses through a vascular region over time. From the IRF, we derive three critical QA parameters Peak Height ($PH_{IRF}$), Area Under the Curve ($AUC_{IRF}$) and Mean Transit Time (MTT) respectively:

$$PH_{IRF} = max(IRF) \tag{9}$$

$$AUC_{IRF} = \int_0^\infty IRF(t)dt \tag{10}$$

$$MTT = \frac{AUC_{IRF}}{PH_{IRF}} \tag{11}$$

$PH_{IRF}$ captures the maximum concentration of contrast media within the ROI during the imaging sequence. $PH_{IRF}$ is needed for understanding the peak blood flow capacity and the vascular responsiveness of the region to the injected contrast. $AUC_{IRF}$ quantifies the total volume of contrast media passing through the region over time. This integral measure reflects the cumulative blood volume that traverses the ROI, providing insights into the overall blood supply to the tissue. MTT represents the average time taken by the contrast media to flow through the specified ROI. MTT is a composite measure that indicates the efficiency of blood flow, with implications for identifying areas of reduced perfusion or delayed transit, which may suggest vascular pathology.

For this study, in the evaluation of hemodynamic characteristics using SVD methodologies, it is essential to distinguish the QA parameters derived from the IRF—$PH_{IRF}$, $AUC_{IRF}$, and MTT—from conventional parameters associated with CT perfusion, such as cerebral blood volume (CBV) and cerebral blood flow (CBF). Given the inherent limitations of DSA, which involves the 2D projection of a 3D object, the full spatial distribution of contrast within the vascular network cannot be precisely determined. This methodological constraint precludes the direct calculation of traditional CTP-like parameters, necessitating an alternative representation of vascular dynamics through IRF-derived metrics for a specific contrast injection and projection angle. Therefore, these QA parameters should be regarded as surrogate measures, providing valuable insights into blood flow characteristics without directly equating them to CBV or CBF.

**2.2 Synthesis of virtual angiogram**

Data analysis was approved by **University at Buffalo** institutional review board. The virtual angiograms were generated using three specific models of ICA. The flow in these models was simulated using Computational Fluid Dynamics (CFD) for inlet velocity 0.25, 0.35 and 0.45m/s with each having injection duration of 0.5, 1.0, 1.5 and 2.0 s, respectively[21]. The segmentation of the models began with the Vitrea 3D station (Vital Images, Inc., Minnetonka, MN), where the focus was on the aneurysm dome and the inlet ROI. Subsequent refinements were applied to the exported stereolithographic (STL) files using Autodesk Meshmixer (Autodesk Inc., San Francisco, CA) to enhance the smoothness of the vessel structure. To create high-resolution meshes that accurately represent the vessel structure, the refined STL files underwent mesh generation using ICEM (Ansys Inc., Canonsburg, PA).

The meshed geometries are exported to Fluent (Ansys Inc., Canonsburg, PA), for steady-state laminar flow analysis. The simulations are based on the incompressible Navier-Stokes equations, with the blood medium modeled as a Newtonian fluid of density (1060 kg/m³) and viscosity (0.0035 Pa·s). For boundary conditions, a parabolic velocity function is defined at each inlet cross-section, representing three different cases with mean velocities of 0.25, 0.35, and 0.45 m/s. The simulation convergence criteria is set to 1e−6 for continuity, employing the SIMPLE scheme with a second-order formulation. Once the velocity field converges, contrast flow is introduced using a passive scalar approach. This involves simulating the injection of a contrast medium at each inlet, with the transport of scalar species through the vessel domain modeled using transient advection-diffusion equations. To initiate the contrast flow, the tracer mass fraction is initialized to zero across the cross-sectional area of the inlet and set to one for the user-defined injection duration. The simulations are designed to meet the convergence criteria of 1e-6 at each timestep, and the SIMPLE scheme is implemented throughout the computational process.

A time-resolved, volumetric contrast agent concentration throughout the in-silico model, is obtained as the solution of the advection-diffusion equations. Each dataset is represented as a non-uniformly spaced, point cloud, where we used MATLAB (MATLAB vR2021b, Natick, MA) for all point cloud operations. Sequential rotations were applied to optimize the viewing angle of the aneurysm. To create a projection sequence, the point cloud data was interpolated onto a structured Cartesian grid, with a maximum grid dimension of 512 voxels, and the other two dimensions scaled accordingly for each model. A binary masking operation was performed to ensure the contrast data interpolation occurs solely within the vessel lumen. This resulted in a gridded virtual angiogram which represents contrast concentration during the pulse period, with each voxel containing a numeric value between zero (no contrast) and one (maximum concentration). The virtual angiogram contains the contrast-enhanced vessel, eliminating other attenuating structures for a clear and precise representation of the contrast distribution.

## 2.3 Data analysis

Analysis in this study was conducted using Python (version 3.10, Wilmington, DE). The process involved creating time-density curves (TDCs) for both aneurysm dome and inlet regions of interest (ROIs) from gridded virtual angiograms. The inlet TDCs specifically were extracted from ROIs located two centimeters proximal to the aneurysm neck, chosen for their reduced tortuosity and foreshortening to ensure accurate flow measurement. We applied three different SVD methods—sSVD with Tikhonov regularization, bSVD, and oSVD—to analyze the QA parameters from each angiogram. These methods were chosen to investigate changes in QA parameters across varying contrast injection durations and flow velocities.

Our analysis was structured to test three objectives:

1. Examination of QA parameter stability across contrast injection durations: This objective involves investigating whether the QA parameters ($PH_{IRF}$, $AUC_{IRF}$, MTT) remain consistent despite variations in the duration of contrast injection. The aim is to assess the sensitivity of these parameters to changes in injection timing, thereby understanding their reliability in different clinical scenarios.
2. Evaluation of SVD methods for QA parameter consistency: Here, the focus is on comparing the effectiveness of different SVD methods (sSVD with Tikhonov regularization, bSVD, and oSVD) in maintaining the consistency of QA parameters. The goal is to identify which method provides the most stable parameter values across varied injection durations, offering insights into the optimal approach for clinical analysis.
3. Analysis of potential bias through slope examination: This objective seeks to identify potential biases in the relationship between QA parameters and injection duration by analyzing their slopes. By understanding how these parameters change with varying injection times, the study aims to develop strategies for mitigating any identified biases, ensuring more accurate and reliable measurements.

Statistical analysis played a key role in this process. Pearson correlation coefficients were used to measure the linear relationship between QA parameters and injection durations for each inlet velocity. This helped determine if the QA parameters significantly depended on the duration of contrast injection. Additionally, root mean square error (RMSE) values were computed to assess the accuracy of our method. The data were summarized by calculating the mean and standard deviation for each QA parameter set, providing a clear view of the data distribution. Scatter plots were used to illustrate how the application of SVD methods affected the relationship between injection duration and QA parameters, with a focus on demonstrating the reduction in injection duration impact.

## 3. Results

We performed SVD variants (sSVD, bSVD and oSVD) on three different virtually generated angiograms. ROIs were drawn over the aneurysm dome and arterial inlet manually to get the QA parameters. Figure 2 shows the virtual angiogram sequences showing the flow of contrast medium at every time step. Figure 3 shows the workflow of the study where the TDC for the aneurysm dome and inlet is extracted from the angiogram. We implement the respective SVD variants to calculate the IRF and we convolve the IRF with inlet function $C_a$ to get $Q_{new}$, which overlaps with the original time density curve of the aneurysm dome, Q in equation 4.

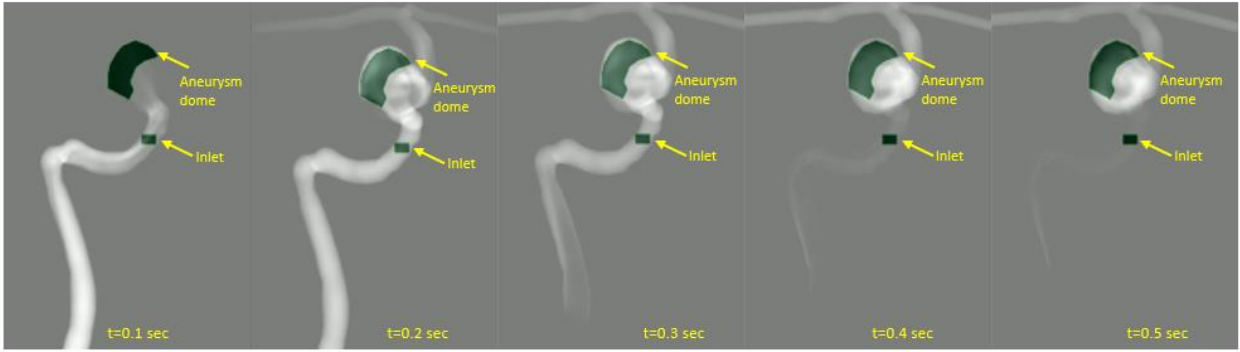

**Figure 2.** Virtual angiogram sequences generated using Computational Fluid Dynamics (CFD) showing the flow of contrast medium through the vasculature.

We performed an analysis of different SVD variants to establish them as normalization methods to eliminate dependence on injection duration. We have taken into consideration three virtual angiogram models with three inlet velocities at 0.25m/s, 0.35m/s and 0.45m/s, and each having an injection duration of 0.5s, 1s, 1.5s and 2s. Table 1 demonstrates Pearson Correlation Coefficient ($P_{coeff}$) and Average of MTT for different velocity and injection duration for a virtually generated angiogram, using no implementation of SVD and implementation of sSVD, bSVD and oSVD. Table 2 in Appendix, demonstrates $PH_{IRF}$, $AUC_{IRF}$ and mean transit time after implementation of SVD ($MTT_{SVD}$) for different velocity and injection duration for a virtually generated angiogram using sSVD, bSVD and oSVD. $SLOPE_{SVD}$ is the slope of $MTT_{SVD}$ for 0.25, 0.35 and 0.45m/sec and SLOPE is the slope of mean transit time before implementation of SVD (MTT) for 0.25, 0.35 and 0.45m/sec. $DIFF_{SL}$ is the difference between $SLOPE_{SVD}$ and SLOPE. $AVG_{SL}$ is the average of the $DIFF_{SL}$ for each SVD variant. AVU is arbitrary volume unit. Table 3 and Table 4 in Appendix, show the optimal truncation for each model and oscillation index for oSVD, respectively.

Figure 4 gives the comparison of slope of MTT before and after the implementation of SVD variants for each inlet velocity at 0.25, 0.35 and 0.45m/sec and Figure 5 summarizes the data shown in Figure 4 using a bar plot.

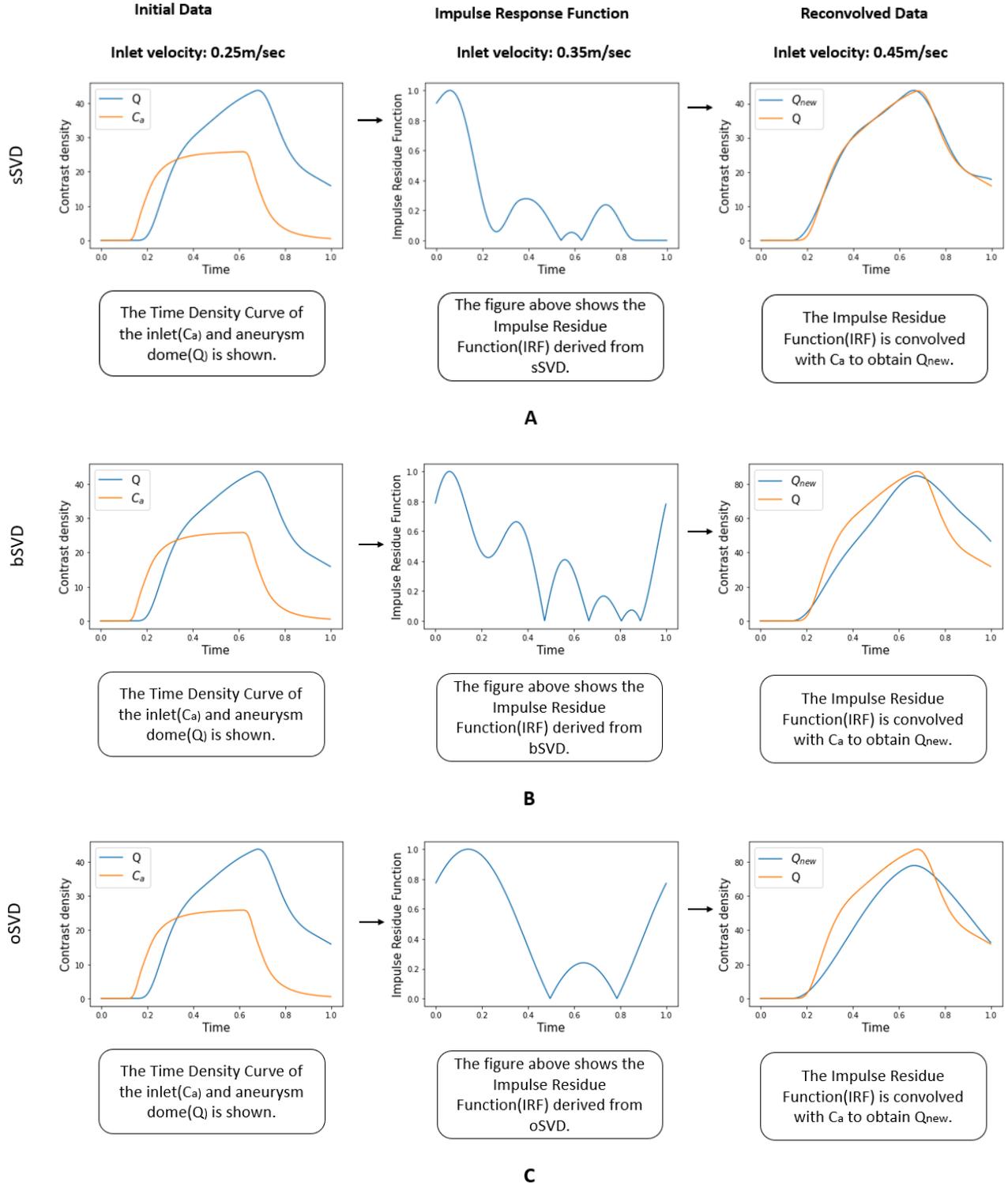

**Figure 3.** The implementation of sSVD (A), bSVD (B) and oSVD (C) is shown. We used virtual angiograms shown in Figure 4 to generate Time Density Curve (TDC) for the inlet and aneurysm dome, and implement sSVD, bSVD and oSVD. Q is the time density curve of the aneurysm dome; $C_a$ is the time density curve of the inlet and $Q_{new}$ is the time density curve obtained by convolving impulse residue function (IRF) with $C_a$. We evaluated IRF, to extract Computed tomography (CT) perfusion parameters: $PH_{IRF}$, $AUC_{IRF}$ and MTT. The leftmost column shows the original TDC of the aneurysm dome and inlet. The second column shows the IRF. The third column shows the generated $Q_{new}$, formed by convolving the inlet function with the IRF. The figures are generated from M1 model.

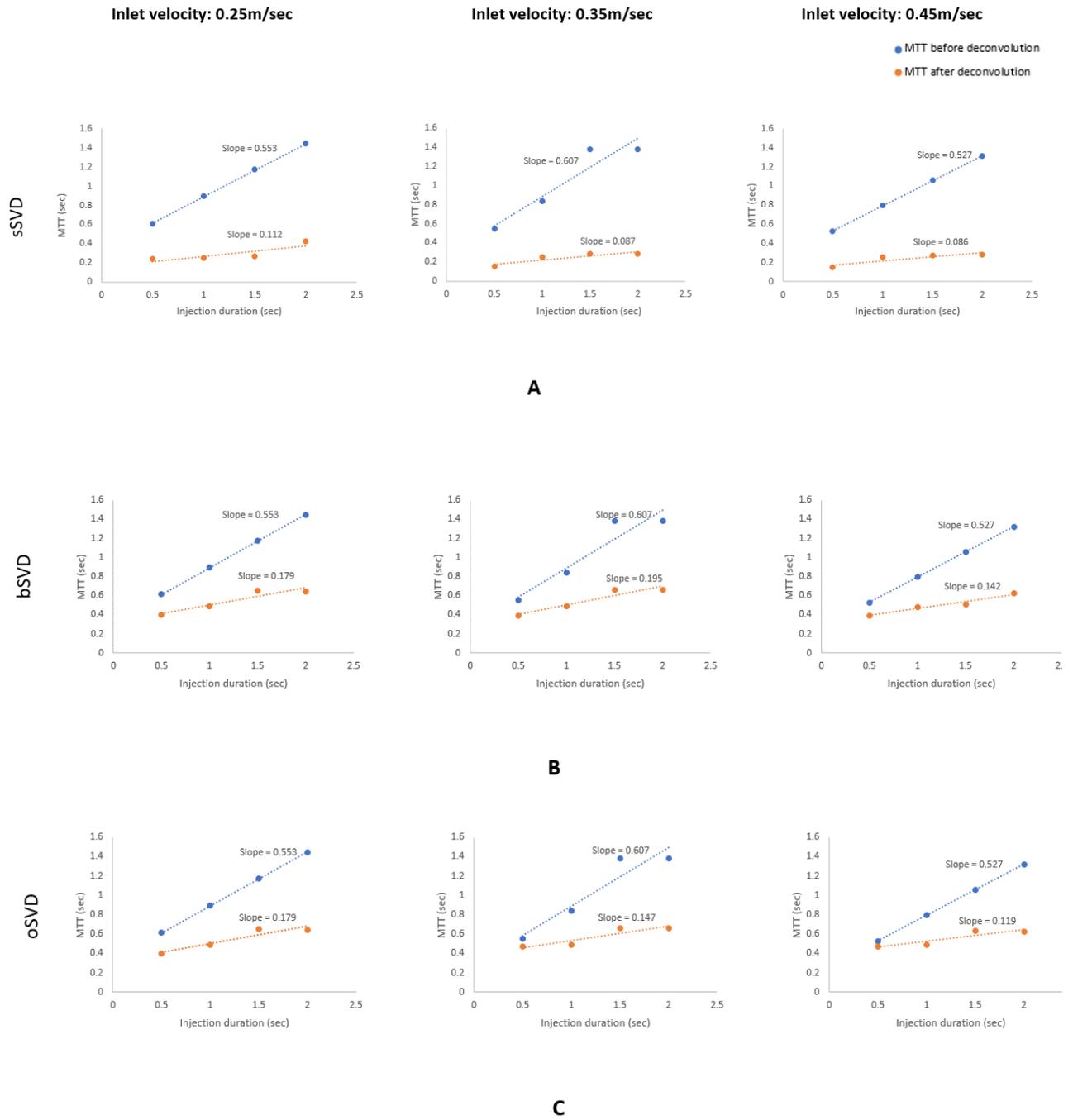

**Figure 4.** Plot of MTT after the implementation of sSVD (A), bSVD (B) and oSVD (C) is shown, at inlet velocity 0.25m/sec, 0.35m/sec and 0.45m/sec. The plots are generated from M1 model.

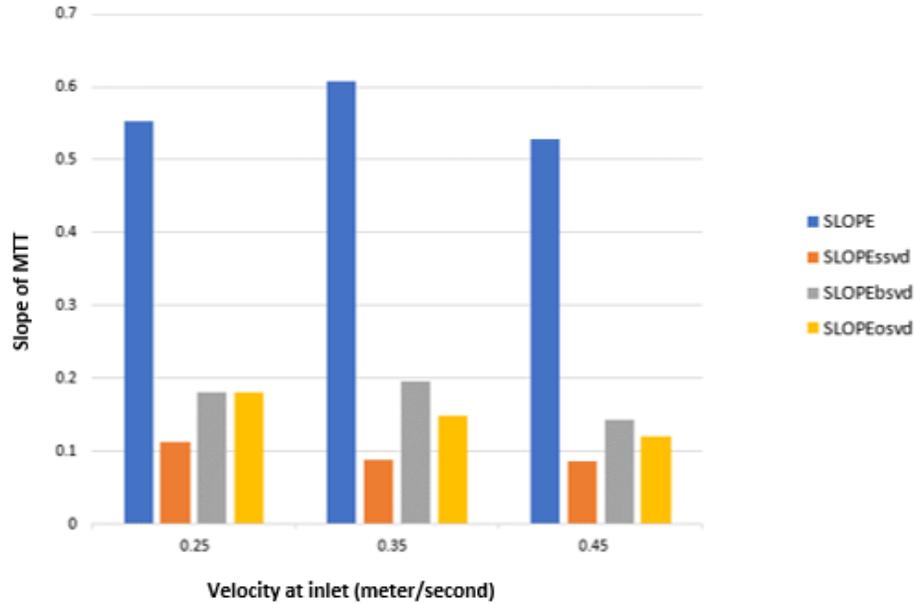

**Figure 5.** Bar plot showing the slope of MTT before and after the implementation of SVD variants: sSVD, bSVD and oSVD for three different velocity profiles at 0.25m/s, 0.35m/s and 0.45m/s, of a specific virtual angiogram. SLOPE demonstrates the slope of MTT before implementation of any SVD variant. $SLOPE_{ssvd}$ gives the slope of MTT after implementation of sSVD. $SLOPE_{bsvd}$ gives the slope of MTT after implementation of bSVD and $SLOPE_{osvd}$ gives the slope of MTT after implementation of oSVD. The plot is generated from M1 model.

Table 2 in Appendix shows the implementation of different SVD variants. We observe a major reduction in slope after the implementation of SVD, which satisfies the hypothesis that implementation of SVD minimizes the impact of hand injection variability on QA parameters. The average difference between the slope of MTT before and after the implementation of SVD for M1 model, using sSVD is 0.467, bSVD is 0.390 and oSVD is 0.413, showing sSVD yields better performance among all the variants. This is well depicted in Figure 5.

The range of $PH_{IRF}$, $AUC_{IRF}$ and MTT for M1 model, using sSVD are $0.011 \pm 0.002$, $0.002 \pm 0.0004$ and $0.260 \pm 0.066$ respectively, for bSVD are $0.006 \pm 0.0009$, $0.003 \pm 0.0004$ and $0.533 \pm 0.104$ respectively, and for oSVD are $0.005 \pm 0.0006$, $0.003 \pm 0.0005$ and $0.557 \pm 0.092$, respectively. We analyzed the data further, to check if there is a correlation between the measured QA values and injection duration. We implemented a Pearson coefficient correlation as a metric, where for M1 model, we found the MTT after the implementation of sSVD shows a correlation of 0.802 with the injection durations. Similarly, for bSVD a correlation of 0.920 was observed, and for oSVD the correlation between MTT and injection durations was observed to be 0.904. Table 1 shows the $P_{coeff}$ for all the three aneurysm models.

12**Table 1:** Demonstrates Pearson Correlation Coefficient ($P_{coeff}$) and Average of MTT for different velocity and injection duration for a virtually generated angiogram using no SVD, sSVD, bSVD and oSVD. M1, M2 and M3 are three aneurysm cases chosen for this study.

| Aneurysm Models: | | M1 | | M2 | | M3 | |
|---|---|---|---|---|---|---|---|
| | | Pcoeff | Average | Pcoeff | Average | Pcoeff | Average |
| SVD variants | No SVD | 0.964 | 0.964 ± 0 | 0.736 | 0.736 ± 0 | 0.701 | 0.701 ± 0 |
| | sSVD | 0.802 | | 0.394 | | 0.276 | |
| | bSVD | 0.920 | 0.876 ± 0.052 | 0.432 | 0.551 ± 0.195 | 0.292 | 0.219 ± 0.092 |
| | oSVD | 0.904 | | 0.827 | | 0.089 | |

## 4. Discussion

Our study delves into the efficacy of Singular Value Decomposition (SVD) variants—namely, standard SVD (sSVD), block-circulant SVD (bSVD), and oscillation index SVD (oSVD)—in standardizing Quantitative Angiography (QA) parameters, despite the variability in contrast injection durations. This thorough exploration into injection variability removal within Digital Subtraction Angiography (DSA) analysis reveals a significant insight: sSVD excels beyond its counterparts, bSVD and oSVD, in minimizing the variability of Mean Transit Time (MTT) metrics. These results also reveal the strengths and weaknesses of each SVD method in addressing the challenges introduced by injection duration variability in DSA analysis.

Previous studies indicate that the MTT values were found to be increase steadily, when increasing the duration of contrast injection, throughout the ROI. However, the MTT of contrast should remain the same since the underlying hemodynamic conditions are unchanged. $PH_{IRF}$ and $AUC_{IRF}$ should also remain constant as the injection duration increases, due to the fact that volume of blood remaining in a particular region and the overall rate at which blood flows are not changing. Similarly, all the other time and intensity related parameters should also remain constant as the injection duration increases. Thus, for the proper implementation of QA parameters for the diagnosis and treatment in clinical setting, the dependence on injection has to be reduced. Thus, in this study we aimed to investigate different SVD variants, to reduce the dependence of injection duration on QA parameters.

We have implemented the SVD variants to evaluate the QA parameters and conducted a statistical analysis on them. Table 2 shows bSVD and oSVD are not as successful as sSVD in reducing the slope of MTT, thus indicating not entirely successful in removing the impact of injection duration variability on QA parameters. This is due to the fact that we use noiseless data in the form of virtual angiograms, whereas bSVD and oSVD algorithms are employed for noisy data.

Another significant aspect of SVD is truncation factor. It is a technique used to reduce dimensionality of the SVD results. It involves retaining only a subset of the most significant singular values and their respective singular vectors while discarding the less significant ones. Truncation thus allows us to focus on the most significant components of the data, which helps in analysis while preserving essential information and removing inconsequential data. For our study, we observed that the optimal truncation, $SVD_{trunc}$, changes for every model, inlet velocity and injection duration, respectively. Table 3 depicts $SVD_{trunc}$ for each combination of SVD variant, inlet velocity and contrast injection duration. Root mean square error (RMSE) quantifies how much $Q_{new}$ and Q deviate from each other, taking into account their magnitude. We observe that on using the mentioned $SVD_{trunc}$ as depicted in the Table 3, we get the least RMSE value between the $Q_{new}$ and Q. RMSE value indicates that on using optimal truncation, we get most of the curve back, after implementation of SVD.

13Similarly, Table 4 demonstrates another significant metric called OI, for the analysis of QA parameters. Table 4 gives the OI obtained during implementation of oSVD for each model. OI quantifies the oscillations at the tail of the IRF. It should be minimized to simplify the analysis of the results. It is evident that on using the depicted $SVD_{trunc}$ we get the minimum OI for the virtual angiograms, after the implementation of oSVD.

The Pearson coefficient correlation values from our study indicate a persistent dependency of QA parameters on injection duration, even after applying Singular Value Decomposition (SVD). Despite expectations that SVD would mitigate this dependency, evidenced by a hoped-for reduction in Pearson correlation, the analysis reveals a sustained high correlation. This limitation stems from the foreshortening error, intrinsic to the 2D imaging system, which distorts the perceived length and shape of anatomical structures due to their alignment relative to the image plane. To counteract this, three-dimensional renderings from CT angiography data could offer a more faithful representation, suggesting that technical adjustments for projection correction might eliminate QA parameters' dependency on injection variability. Nonetheless, our results demonstrate a notable reduction in this dependency, as quantified by the changes in metric slopes for sSVD, bSVD, and oSVD, detailed in Table 2, with average differences of 0.467, 0.390, and 0.413, respectively. This indicates progress towards diminishing the influence of injection duration on QA metrics, albeit within the confines of the study's methodological constraints.

There are some limitations to our study. This study, conducted in silico, presents an idealized scenario devoid of the common challenges encountered in clinical settings. Notably, our simulations did not incorporate noise or account for clinical factors such as patient motion or X-ray imaging-related artifacts, which can significantly affect the quality and interpretability of angiographic images. Additionally, the investigation was limited to three aneurysm cases, constraining the breadth of our findings and their applicability across the diverse spectrum of intracranial aneurysms. Furthermore, our approach to simulating the angiography process did not employ a two-fluid model, which might have offered a more nuanced representation of the interaction between contrast media and blood flow within the vascular system. The simulated injections were conducted under relatively ideal conditions, with fluid particles labeled for the duration of the injection to ensure pristine data in terms of temporal and spatial resolution. We also observe the MTT obtained from sSVD is much lower when compared to the MTT obtained from bSVD and oSVD methods. We have observed similar trends in previous studies as well, where sSVD shows lower MTT values as compared to bSVD or oSVD methods[22,23]. We believe this is due to the formulation of Toeplitz matrix in sSVD, while, block-circulant matrix in bSVD and oSVD methods, which is the cause of this striking difference. This methodological choice, while beneficial for isolating the effects of SVD algorithms on QA parameter variability, may not fully capture the complexity of real-world angiographic procedures and the dynamic nature of blood flow and contrast dispersion. Future research should aim to incorporate these real-world factors and expand the scope of study to include a wider array of aneurysm types and conditions, potentially employing more sophisticated fluid dynamics models to enhance the realism and applicability of the results.

## 5. Conclusions

Implementation of SVD deconvolution on QA parameters can be used as a normalization method to reduce dependence of QA parameters on injection duration. This correction of QA parameters helps in the correct diagnosis of neurovascular diseases.

We successfully implemented SVD deconvolution in this study and determined its effectiveness in reducing the impact of injection duration on QA parameters. We have effectively established SVD as a normalization method in preventing QA parameters from being skewed.

## 6. Acknowledgements

This work was supported by QAS.AI and NSF STTR Award # 2111865## 7. Conflict of interest Statement

The authors have no relevant conflicts of interest to disclose.

## 8. References

1. Jersey AM, Foster DM. Cerebral Aneurysm. In: *StatPearls.* Treasure Island (FL)2023.
2. Hashimoto T, Meng H, Young WL. Intracranial aneurysms: links among inflammation, hemodynamics and vascular remodeling. *Neurol Res.* 2006;28(4):372-380.
3. Shakur SF, Brunozzi D, Hussein AE, et al. Validation of cerebral arteriovenous malformation hemodynamics assessed by DSA using quantitative magnetic resonance angiography: preliminary study [published online ahead of print 20170224]. *J Neurointerv Surg.* 2018;10(2):156-161.
4. Metzmann U, Lindner P, Thelen M. [Quantitative determination of cerebral perfusion using digital subtraction angiography (DSA)]. *Rofo.* 1990;153(1):41-47.
5. Mondal P, Udin M, Shiraz Bhurwani MM, Williams K, Ionita C. *Intra-operative optimal flow diverter selection for intracranial aneurysm treatment using angiographic parametric imaging: feasibility study.* Vol 12468: SPIE; 2023.
6. Okamoto K, Ito J, Sakai K, Yoshimura S. The principle of digital subtraction angiography and radiological protection. *Interv Neuroradiol.* 2000;6:25-31.
7. Rava R, Allman A, Rudin S, Ionita C. *Effect of truncated singular value decomposition on digital subtraction angiography derived angiographic parametric imaging maps.* Vol 11312: SPIE; 2020.
8. Ionita CN, Garcia VL, Bednarek DR, et al. Effect of injection technique on temporal parametric imaging derived from digital subtraction angiography in patient specific phantoms. *Proc SPIE Int Soc Opt Eng.* 2014;9038:90380L.
9. White R, Shields A, Nagesh SV, et al. Investigating Angiographic Injection Parameters for Cerebral Aneurysm Hemodynamic Characterization Using Patient-Specific Simulated Angiograms [published online ahead of print 20230410]. *Proc SPIE Int Soc Opt Eng.* 2023;12468.
10. Smith MR, Lu H, Trochet S, Frayne R. Removing the effect of SVD algorithmic artifacts present in quantitative MR perfusion studies. *Magn Reson Med.* 2004;51(3):631-634.
11. Salluzzi M, Frayne R, Smith MR. An alternative viewpoint of the similarities and differences of SVD and FT deconvolution algorithms used for quantitative MR perfusion studies. *Magn Reson Imaging.* 2005;23(3):481-492.
12. Morishita T, Tanabe N, Masuo O, et al. [Comparison of Bayesian Estimation and SVD Methods for CT Perfusion in Patients with Acute Stroke] [published online ahead of print 20230214]. *Nihon Hoshasen Gijutsu Gakkai Zasshi.* 2023;79(4):307-312.
13. Li J. The Fick principle remains accurate to calculate cardiac output under hyperoxia. *Acta Anaesth Scand.* 2019;63(6):827-827.
14. Schmidt D. A Survey of Singular Value Decomposition Methods for Distributed Tall/Skinny Data. *Proceedings of Scala 2020: 11th Workshop on Latest Advances in Scalable Algorithms for Large-Scale Systems.* 2020. doi: 10.1109/ScalA51936.2020.00009:27-34.
15. Haldrup K. Singular value decomposition as a tool for background corrections in time-resolved XFEL scattering data. *Philos T R Soc B.* 2014;369(1647).

# Appendix

**Table 2:** Demonstrates $PH_{IRF}$, $AUC_{IRF}$ and $MTT_{SVD}$ for different velocity and injection duration for a virtually generated angiogram using sSVD, bSVD and oSVD. $MTT_{SVD}$ is the mean transit time evaluated after the implementation of SVD and MTT is the mean transit time before the implementation of SVD. $SLOPE_{SVD}$ is the slope of $MTT_{SVD}$ for 0.25, 0.35 and 0.45m/sec and SLOPE is the slope of MTT for 0.25, 0.35 and 0.45m/sec. $DIFF_{SL}$ is the difference between $SLOPE_{SVD}$ and SLOPE. $AVG_{SL}$ is the average of the $DIFF_{SL}$ for each SVD variant. AVU is arbitrary volume unit. A. M1 model B. M2 model C. M3 model

| SVD variant | Velocity at inlet (m/sec) | Injection duration (sec) | $PH_{IRF}$ (AVU/min) | $AUC_{IRF}$ (AVU) | $MTT_{SVD}$ (sec) | MTT (sec) | $Slope_{SVD}$ | Slope | $DIFF_{SL}$ | $AVG_{SL}$ |
|---|---|---|---|---|---|---|---|---|---|---|
| sSVD | 0.25 | 0.5 | 0.009 | 0.002 | 0.240 | 0.611 | 0.112 | 0.553 | 0.440 | 0.467 |
| | | 1 | 0.010 | 0.002 | 0.247 | 0.895 | | | | |
| | | 1.5 | 0.010 | 0.002 | 0.266 | 1.171 | | | | |
| | | 2 | 0.006 | 0.002 | 0.421 | 1.442 | | | | |
| | 0.35 | 0.5 | 0.014 | 0.002 | 0.154 | 0.552 | 0.087 | 0.607 | 0.520 | |
| | | 1 | 0.011 | 0.002 | 0.251 | 0.836 | | | | |
| | | 1.5 | 0.010 | 0.003 | 0.288 | 1.382 | | | | |
| | | 2 | 0.010 | 0.003 | 0.288 | 1.382 | | | | |
| | 0.45 | 0.5 | 0.017 | 0.002 | 0.147 | 0.525 | 0.086 | 0.527 | 0.441 | |
| | | 1 | 0.012 | 0.003 | 0.258 | 0.797 | | | | |
| | | 1.5 | 0.011 | 0.003 | 0.276 | 1.059 | | | | |
| | | 2 | 0.011 | 0.003 | 0.285 | 1.317 | | | | |
| bSVD | 0.25 | 0.5 | 0.006 | 0.002 | 0.400 | 0.611 | 0.179 | 0.553 | 0.373 | 0.390 |
| | | 1 | 0.006 | 0.002 | 0.491 | 0.895 | | | | |
| | | 1.5 | 0.004 | 0.003 | 0.653 | 1.171 | | | | |
| | | 2 | 0.005 | 0.003 | 0.646 | 1.442 | | | | |
| | 0.35 | 0.5 | 0.007 | 0.002 | 0.390 | 0.552 | 0.195 | 0.607 | 0.411 | |
| | | 1 | 0.005 | 0.002 | 0.490 | 0.836 | | | | |
| | | 1.5 | 0.005 | 0.003 | 0.660 | 1.382 | | | | |
| | | 2 | 0.005 | 0.003 | 0.660 | 1.382 | | | | |
| | 0.45 | 0.5 | 0.007 | 0.003 | 0.393 | 0.525 | 0.142 | 0.527 | 0.385 | |
| | | 1 | 0.007 | 0.003 | 0.480 | 0.797 | | | | |
| | | 1.5 | 0.006 | 0.003 | 0.506 | 1.059 | | | | |
| | | 2 | 0.005 | 0.003 | 0.621 | 1.317 | | | | |
| oSVD | 0.25 | 0.5 | 0.006 | 0.002 | 0.400 | 0.611 | 0.179 | 0.553 | 0.373 | 0.413 |
| | | 1 | 0.006 | 0.002 | 0.491 | 0.895 | | | | |
| | | 1.5 | 0.004 | 0.003 | 0.653 | 1.171 | | | | |
| | | 2 | 0.005 | 0.003 | 0.646 | 1.442 | | | | |
| | 0.35 | 0.5 | 0.004 | 0.002 | 0.470 | 0.552 | 0.147 | 0.607 | 0.459 | |
| | | 1 | 0.005 | 0.002 | 0.490 | 0.836 | | | | |
| | | 1.5 | 0.005 | 0.003 | 0.660 | 1.382 | | | | |
| | | 2 | 0.005 | 0.003 | 0.660 | 1.382 | | | | |
| | 0.45 | 0.5 | 0.005 | 0.002 | 0.471 | 0.525 | 0.119 | 0.527 | 0.407 | |
| | | 1 | 0.006 | 0.002 | 0.488 | 0.797 | | | | |
| | | 1.5 | 0.005 | 0.003 | 0.638 | 1.059 | | | | |
| | | 2 | 0.005 | 0.003 | 0.621 | 1.317 | | | | |

**A**

| SVD variant | Velocity at inlet (m/sec) | Injection duration (sec) | $PH_{IRF}$ (AVU/min) | $AUC_{IRF}$ (AVU) | $MTT_{SVD}$ (sec) | MTT (sec) | $Slope_{SVD}$ | Slope | $DIFF_{SL}$ | $AVG_{SL}$ |
|---|---|---|---|---|---|---|---|---|---|---|
| sSVD | 0.25 | 0.5 | 0.007 | 0.001 | 0.191 | 0.647 | 0.094 | 0.516 | 0.422 | 0.295 |
| | | 1 | 0.005 | 0.001 | 0.307 | 0.903 | | | | |
| | | 1.5 | 0.005 | 0.002 | 0.331 | 1.154 | | | | |
| | | 2 | 0.005 | 0.002 | 0.34 | 1.425 | | | | |
| | 0.35 | 0.5 | 0.008 | 0.001 | 0.188 | 0.55 | 0.038 | 0.244 | 0.206 | |
| | | 1 | 0.006 | 0.002 | 0.248 | 0.678 | | | | |
| | | 1.5 | 0.006 | 0.002 | 0.266 | 0.804 | | | | |
| | | 2 | 0.006 | 0.002 | 0.246 | 0.914 | | | | |
| | 0.45 | 0.5 | 0.009 | 0.002 | 0.183 | 0.484 | -0.008 | 0.249 | 0.257 | |
| | | 1 | 0.008 | 0.002 | 0.197 | 0.609 | | | | |
| | | 1.5 | 0.008 | 0.001 | 0.177 | 0.731 | | | | |
| | | 2 | 0.008 | 0.001 | 0.177 | 0.858 | | | | |
| bSVD | 0.25 | 0.5 | 0.005 | 0.002 | 0.35 | 0.647 | 0.11 | 0.516 | 0.406 | 0.268 |
| | | 1 | 0.005 | 0.002 | 0.375 | 0.903 | | | | |
| | | 1.5 | 0.004 | 0.002 | 0.499 | 1.154 | | | | |
| | | 2 | 0.004 | 0.002 | 0.493 | 1.425 | | | | |
| | 0.35 | 0.5 | 0.005 | 0.002 | 0.377 | 0.55 | 0.102 | 0.244 | 0.142 | |
| | | 1 | 0.005 | 0.002 | 0.391 | 0.678 | | | | |
| | | 1.5 | 0.004 | 0.002 | 0.489 | 0.804 | | | | |
| | | 2 | 0.004 | 0.002 | 0.515 | 0.914 | | | | |
| | 0.45 | 0.5 | 0.006 | 0.002 | 0.304 | 0.484 | 0.002 | 0.249 | 0.247 | |
| | | 1 | 0.008 | 0.002 | 0.237 | 0.609 | | | | |
| | | 1.5 | 0.007 | 0.002 | 0.27 | 0.731 | | | | |
| | | 2 | 0.006 | 0.002 | 0.297 | 0.858 | | | | |
| oSVD | 0.25 | 0.5 | 0.004 | 0.002 | 0.389 | 0.647 | 0.208 | 0.516 | 0.309 | 0.141 |
| | | 1 | 0.004 | 0.002 | 0.485 | 0.903 | | | | |
| | | 1.5 | 0.004 | 0.002 | 0.499 | 1.154 | | | | |
| | | 2 | 0.003 | 0.002 | 0.73 | 1.425 | | | | |
| | 0.35 | 0.5 | 0.005 | 0.002 | 0.377 | 0.55 | 0.217 | 0.244 | 0.027 | |
| | | 1 | 0.004 | 0.002 | 0.446 | 0.678 | | | | |
| | | 1.5 | 0.003 | 0.002 | 0.573 | 0.804 | | | | |
| | | 2 | 0.003 | 0.002 | 0.698 | 0.914 | | | | |
| | 0.45 | 0.5 | 0.005 | 0.002 | 0.366 | 0.484 | 0.161 | 0.249 | 0.089 | |
| | | 1 | 0.007 | 0.002 | 0.263 | 0.609 | | | | |
| | | 1.5 | 0.004 | 0.002 | 0.443 | 0.731 | | | | |
| | | 2 | 0.003 | 0.002 | 0.574 | 0.858 | | | | |

**B**

| SVD variant | Velocity at inlet (m/sec) | Injection duration (sec) | $PH_{IRF}$ (AVU/min) | $AUC_{IRF}$ (AVU) | $MTT_{SVD}$ (sec) | MTT (sec) | $Slope_{SVD}$ | Slope | $DIFF_{SL}$ | $AVG_{SL}$ |
|---|---|---|---|---|---|---|---|---|---|---|
| sSVD | 0.25 | 0.5 | 0.010 | 0.002 | 0.206 | 0.671 | 0.073 | 0.145 | 0.071 | 0.147 |
| | | 1 | 0.006 | 0.002 | 0.362 | 0.921 | | | | |
| | | 1.5 | 0.006 | 0.002 | 0.367 | 0.918 | | | | |
| | | 2 | 0.007 | 0.002 | 0.327 | 0.913 | | | | |
| | 0.35 | 0.5 | 0.009 | 0.002 | 0.238 | 0.545 | 0.024 | 0.215 | 0.191 | |
| | | 1 | 0.009 | 0.002 | 0.263 | 0.673 | | | | |
| | | 1.5 | 0.008 | 0.002 | 0.257 | 0.798 | | | | |
| | | 2 | 0.008 | 0.002 | 0.280 | 0.862 | | | | |
| | 0.45 | 0.5 | 0.012 | 0.002 | 0.182 | 0.484 | -0.002 | 0.176 | 0.178 | |
| | | 1 | 0.010 | 0.002 | 0.210 | 0.609 | | | | |
| | | 1.5 | 0.011 | 0.002 | 0.185 | 0.734 | | | | |
| | | 2 | 0.011 | 0.002 | 0.186 | 0.736 | | | | |
| bSVD | 0.25 | 0.5 | 0.009 | 0.002 | 0.234 | 0.671 | 0.004 | 0.145 | 0.14 | 0.13 |
| | | 1 | 0.005 | 0.003 | 0.578 | 0.921 | | | | |
| | | 1.5 | 0.007 | 0.002 | 0.345 | 0.918 | | | | |
| | | 2 | 0.007 | 0.002 | 0.319 | 0.913 | | | | |
| | 0.35 | 0.5 | 0.008 | 0.003 | 0.301 | 0.545 | 0.015 | 0.215 | 0.2 | |
| | | 1 | 0.009 | 0.003 | 0.299 | 0.673 | | | | |
| | | 1.5 | 0.008 | 0.003 | 0.328 | 0.798 | | | | |
| | | 2 | 0.008 | 0.003 | 0.315 | 0.862 | | | | |
| | 0.45 | 0.5 | 0.009 | 0.003 | 0.284 | 0.484 | 0.126 | 0.176 | 0.05 | |
| | | 1 | 0.011 | 0.003 | 0.233 | 0.609 | | | | |
| | | 1.5 | 0.006 | 0.003 | 0.423 | 0.734 | | | | |
| | | 2 | 0.006 | 0.003 | 0.43 | 0.736 | | | | |
| oSVD | 0.25 | 0.5 | 0.012 | 0.002 | 0.168 | 0.671 | 0.034 | 0.145 | 0.11 | 0.166 |
| | | 1 | 0.007 | 0.003 | 0.372 | 0.921 | | | | |
| | | 1.5 | 0.01 | 0.002 | 0.231 | 0.918 | | | | |
| | | 2 | 0.009 | 0.003 | 0.272 | 0.913 | | | | |
| | 0.35 | 0.5 | 0.01 | 0.003 | 0.243 | 0.545 | -0.043 | 0.215 | 0.258 | |
| | | 1 | 0.016 | 0.002 | 0.129 | 0.673 | | | | |
| | | 1.5 | 0.015 | 0.002 | 0.156 | 0.798 | | | | |
| | | 2 | 0.014 | 0.002 | 0.163 | 0.862 | | | | |
| | 0.45 | 0.5 | 0.02 | 0.002 | 0.093 | 0.484 | 0.044 | 0.176 | 0.132 | |
| | | 1 | 0.017 | 0.003 | 0.147 | 0.609 | | | | |
| | | 1.5 | 0.017 | 0.002 | 0.139 | 0.734 | | | | |
| | | 2 | 0.015 | 0.003 | 0.169 | 0.736 | | | | |

**C**

**Table 3:** Demonstrates the optimal truncation (SVD$_{trunc}$) and Root Mean Square Error (RMSE) value between the original TDC of the aneurysm dome (Q) and Q$_{new}$ generated by convolving the inlet function (C$_a$) with the IRF for different velocity and injection duration. A. M1 model B. M2 model C. M3 model

| SVD variant | Velocity at inlet (m/sec) | Injection duration (sec) | SVD$_{trunc}$ | RMSE |
|---|---|---|---|---|
| sSVD | 0.25 | 0.5 | 0.08-0.2 | 0.790 |
| | | 1 | 0.05-0.1 | 0.790 |
| | | 1.5 | 0.05-0.07 | 0.903 |
| | | 2 | 0.04-0.05 | 0.976 |
| | 0.35 | 0.5 | 0.05-0.09 | 0.829 |
| | | 1 | 0.03-0.04 | 0.957 |
| | | 1.5 | 0.04-0.07 | 2.298 |
| | | 2 | 0.04-0.07 | 2.298 |
| | 0.45 | 0.5 | 0.03-0.07 | 0.658 |
| | | 1 | 0.05-0.06 | 1.944 |
| | | 1.5 | 0.02-0.04 | 1.578 |
| | | 2 | 0.01-0.03 | 2.383 |
| bSVD | 0.25 | 0.5 | 0.3-0.4 | 5.816 |
| | | 1 | 0.09-0.1 | 4.406 |
| | | 1.5 | 0.09-0.1 | 6.312 |
| | | 2 | 0.06-0.1 | 7.376 |
| | 0.35 | 0.5 | 0.1-0.4 | 7.652 |
| | | 1 | 0.09-0.1 | 6.399 |
| | | 1.5 | 0.04-0.05 | 7.129 |
| | | 2 | 0.04-0.05 | 7.129 |
| | 0.45 | 0.5 | 0.2-0.4 | 8.937 |
| | | 1 | 0.09-0.1 | 8.396 |
| | | 1.5 | 0.07-0.09 | 7.470 |
| | | 2 | 0.05 | 7.078 |
| oSVD | 0.25 | 0.5 | 0.2-0.4 | 5.816 |
| | | 1 | 0.07-0.1 | 4.406 |
| | | 1.5 | 0.08-0.1 | 6.312 |
| | | 2 | 0.04-0.1 | 7.376 |
| | 0.35 | 0.5 | 0.09-0.4 | 7.652 |
| | | 1 | 0.08-0.1 | 6.399 |
| | | 1.5 | 0.04-0.05 | 7.129 |
| | | 2 | 0.04-0.05 | 7.129 |
| | 0.45 | 0.5 | 0.1-0.4 | 8.937 |
| | | 1 | 0.09-0.1 | 8.396 |
| | | 1.5 | 0.06-0.09 | 7.470 |
| | | 2 | 0.05 | 7.078 |

A

| SVD variant | Velocity at inlet (m/sec) | Injection duration (sec) | SVD$_{trunc}$ | RMSE |
|---|---|---|---|---|
| sSVD | 0.25 | 0.5 | 0.01-0.1 | 1.452 |
| | | 1 | 0.01-0.1 | 1.769 |
| | | 1.5 | 0.01-0.1 | 2.682 |
| | | 2 | 0.04-0.1 | 7.023 |
| | 0.35 | 0.5 | 0.01-0.1 | 3.557 |
| | | 1 | 0.01-0.1 | 3.854 |
| | | 1.5 | 0.01-0.1 | 5.940 |
| | | 2 | 0.01-0.1 | 6.563 |
| | 0.45 | 0.5 | 0.01-0.09 | 1.751 |
| | | 1 | 0.01-0.09 | 6.299 |
| | | 1.5 | 0.01-0.07 | 3.472 |
| | | 2 | 0.03-0.07 | 6.458 |
| bSVD | 0.25 | 0.5 | 0.05-0.1 | 5.374 |
| | | 1 | 0.06-0.1 | 6.025 |
| | | 1.5 | 0.08-0.1 | 7.489 |
| | | 2 | 0.07-0.09 | 9.862 |
| | 0.35 | 0.5 | 0.1-0.4 | 9.471 |
| | | 1 | 0.2-0.4 | 11.552 |
| | | 1.5 | 0.2-0.4 | 13.335 |
| | | 2 | 0.2-0.4 | 14.604 |
| | 0.45 | 0.5 | 0.06-0.3 | 10.591 |
| | | 1 | 0.06-0.1 | 8.179 |
| | | 1.5 | 0.07-0.1 | 9.362 |
| | | 2 | 0.07-0.1 | 11.386 |
| oSVD | 0.25 | 0.5 | 0.05-0.1 | 5.374 |
| | | 1 | 0.06-0.1 | 6.025 |
| | | 1.5 | 0.08-0.1 | 7.489 |
| | | 2 | 0.07-0.09 | 9.862 |
| | 0.35 | 0.5 | 0.1-0.4 | 9.471 |
| | | 1 | 0.2-0.4 | 11.552 |
| | | 1.5 | 0.2-0.4 | 13.335 |
| | | 2 | 0.2-0.4 | 14.604 |
| | 0.45 | 0.5 | 0.06-0.3 | 10.591 |
| | | 1 | 0.06-0.1 | 8.179 |
| | | 1.5 | 0.07-0.1 | 9.362 |
| | | 2 | 0.07-0.1 | 11.386 |

B

| SVD variant | Velocity at inlet (m/sec) | Injection duration (sec) | $SVD_{trunc}$ | RMSE |
|---|---|---|---|---|
| sSVD | 0.25 | 0.5 | 0.01-0.06 | 2.361 |
| | | 1 | 0.08-0.1 | 4.329 |
| | | 1.5 | 0.06-0.1 | 4.359 |
| | | 2 | 0.07-0.1 | 3.882 |
| | 0.35 | 0.5 | 0.01-0.2 | 2.792 |
| | | 1 | 0.07-0.2 | 4.585 |
| | | 1.5 | 0.05-0.1 | 5.903 |
| | | 2 | 0.07-0.1 | 4.617 |
| | 0.45 | 0.5 | 0.07-0.2 | 4.298 |
| | | 1 | 0.08-0.09 | 4.857 |
| | | 1.5 | 0.04-0.1 | 4.867 |
| | | 2 | 0.02-0.06 | 4.894 |
| bSVD | 0.25 | 0.5 | 0.05-0.1 | 2.800 |
| | | 1 | 0.2-0.6 | 9.292 |
| | | 1.5 | 0.05-0.09 | 5.160 |
| | | 2 | 0.04-0.06 | 4.641 |
| | 0.35 | 0.5 | 0.07-0.09 | 8.538 |
| | | 1 | 0.01-0.02 | 3.951 |
| | | 1.5 | 0.05-0.06 | 7.270 |
| | | 2 | 0.01-0.05 | 6.965 |
| | 0.45 | 0.5 | 0.01-0.04 | 1.779 |
| | | 1 | 0.06-0.1 | 6.452 |
| | | 1.5 | 0.01-0.02 | 7.368 |
| | | 2 | 0.01-0.03 | 6.831 |
| oSVD | 0.25 | 0.5 | 0.05-0.1 | 2.800 |
| | | 1 | 0.2-0.6 | 9.292 |
| | | 1.5 | 0.05-0.09 | 5.160 |
| | | 2 | 0.04-0.06 | 4.641 |
| | 0.35 | 0.5 | 0.07-0.09 | 8.538 |
| | | 1 | 0.01-0.02 | 3.951 |
| | | 1.5 | 0.05-0.06 | 7.270 |
| | | 2 | 0.01-0.05 | 6.965 |
| | 0.45 | 0.5 | 0.01-0.04 | 1.779 |
| | | 1 | 0.06-0.1 | 6.452 |
| | | 1.5 | 0.01-0.02 | 7.368 |
| | | 2 | 0.01-0.03 | 6.831 |

C

**Table 4:** Demonstrates the oscillation index (OI) evaluated for the optimal truncation (SVD$_{trunc}$), after the implementation of oscillation index single value decomposition(oSVD). A. M1 model B. M2 model C. M3 model

| SVD variant | Velocity at inlet (m/sec) | Injection duration (sec) | SVDtrunc | RMS | OI |
|---|---|---|---|---|---|
| oSVD | 0.25 | 0.5 | 0.2-0.4 | 5.816 | 0.027 |
| | | 1 | 0.07-0.1 | 4.406 | 0.029 |
| | | 1.5 | 0.08-0.1 | 6.312 | 0.015 |
| | | 2 | 0.04-0.1 | 7.376 | 0.015 |
| | 0.35 | 0.5 | 0.09-0.4 | 7.652 | 0.024 |
| | | 1 | 0.08-0.1 | 6.399 | 0.024 |
| | | 1.5 | 0.04-0.05 | 7.129 | 0.019 |
| | | 2 | 0.04-0.05 | 7.129 | 0.019 |
| | 0.45 | 0.5 | 0.1-0.4 | 8.937 | 0.023 |
| | | 1 | 0.09-0.1 | 8.396 | 0.020 |
| | | 1.5 | 0.06-0.09 | 7.470 | 0.018 |
| | | 2 | 0.05 | 7.078 | 0.016 |

A

| SVD variant | Velocity at inlet (m/sec) | Injection duration (sec) | SVDtrunc | RMS | OI |
|---|---|---|---|---|---|
| oSVD | 0.25 | 0.5 | 0.05-0.1 | 5.374 | 0.081 |
| | | 1 | 0.06-0.1 | 6.025 | 0.031 |
| | | 1.5 | 0.08-0.1 | 7.489 | 0.018 |
| | | 2 | 0.07-0.09 | 9.862 | 0.012 |
| | 0.35 | 0.5 | 0.1-0.4 | 9.471 | 0.095 |
| | | 1 | 0.2-0.4 | 11.552 | 0.013 |
| | | 1.5 | 0.2-0.4 | 13.355 | 0.007 |
| | | 2 | 0.2-0.4 | 14.604 | 0.005 |
| | 0.45 | 0.5 | 0.06-0.3 | 10.591 | 0.030 |
| | | 1 | 0.06-0.1 | 8.179 | 0.092 |
| | | 1.5 | 0.07-0.1 | 9.362 | 0.038 |
| | | 2 | 0.07-0.1 | 11.386 | 0.023 |

B

| SVD variant | Velocity at inlet (m/sec) | Injection duration (sec) | SVDtrunc | RMS | OI |
|---|---|---|---|---|---|
| oSVD | 0.25 | 0.5 | 0.05-0.1 | 2.800 | 0.092 |
| | | 1 | 0.2-0.6 | 9.292 | 0.007 |
| | | 1.5 | 0.05-0.09 | 5.160 | 0.046 |
| | | 2 | 0.04-0.06 | 4.641 | 0.05 |
| | 0.35 | 0.5 | 0.07-0.09 | 8.538 | 0.106 |
| | | 1 | 0.01-0.02 | 3.951 | 0.172 |
| | | 1.5 | 0.05-0.06 | 7.270 | 0.06 |
| | | 2 | 0.01-0.05 | 6.965 | 0.132 |
| | 0.45 | 0.5 | 0.01-0.04 | 1.779 | 0.264 |
| | | 1 | 0.06-0.1 | 6.452 | 0.093 |
| | | 1.5 | 0.01-0.02 | 7.368 | 0.195 |
| | | 2 | 0.01-0.03 | 6.831 | 0.112 |

C